\documentclass{ws-procs961x669}            
\usepackage{ws-toc,ws-multind}

\begin{document}



\cleardoublepage





\title{Tracing cosmic stretch marks: probing scale invariance in the early Universe}
\author{Mariaveronica De Angelis$^*$}

\address{School of Mathematical and Physical sciences, University of Sheffield, Hounsfield Road, Sheffield S3
7RH, United Kingdom, \\$^*$Speaker. E-mail: mdeangelis1@sheffield.ac.uk}

\author{Chiara Cecchini}
\index{author}{Author, F.} 

\address{Department of Physics, University of Trento, Via Sommarive 14, 38122 Povo (TN), Italy\\ Trento Institute for Fundamental Physics and Applications (TIFPA)-INFN, Via Sommarive
14, 38122 Povo (TN), Italy}



\author{Massimiliano Rinaldi}

\address{Department of Physics, University of Trento, Via Sommarive 14, 38122 Povo (TN), Italy\\ Trento Institute for Fundamental Physics and Applications (TIFPA)-INFN, Via Sommarive
14, 38122 Povo (TN), Italy}


\begin{abstract}
This paper investigates a scale-invariant inflationary model characterized by a scalar field non-minimally coupled to gravity and a curvature term quadratic in the Ricci scalar. The model’s dynamic is analyzed using a full numerical solution of the two-field system, going beyond previous analytical studies. We derive robust constraints on the model parameters using the latest Cosmic Microwave Background (CMB) data from Planck and BICEP/Keck. The study confirms that scale-invariance effectively reduces the system to single-field dynamics, eliminating entropy perturbations and ensuring stability. Key predictions include a minimal level of primordial gravitational waves with a tensor-to-scalar ratio 
$r \geq 0.003$, which upcoming CMB experiments are well-positioned to test. The model is compared to Starobinsky and $\alpha -$attractor inflation, with future observations of tensor modes offering a potential discriminator between them. Overall, the results suggest that scale-invariant inflation is a viable and competitive framework for explaining early universe dynamics and predicting cosmological observables.
\end{abstract}


\section{Introduction}
Cosmic inflation, a period of accelerated expansion in the early universe, has become a cornerstone of modern cosmology. Initially proposed to resolve the flatness, horizon, and monopole problems of the Big Bang model, inflation also provides a compelling explanation for the observed structure of the universe. Tiny quantum fluctuations, stretched to cosmic scales during inflation, are thought to seed the large-scale structure of the universe and the anisotropies in the Cosmic Microwave Background (CMB). However, despite its observational success, the theoretical underpinnings of inflation remain a subject of debate, with many competing models vying to describe this early phase of cosmic evolution.

Most inflationary models rely on a scalar field, called the inflaton, rolling slowly down a potential, driving an exponential expansion of space. Single-field inflation models, where a single scalar field drives inflation, have been extensively studied. These models predict a spectrum of primordial perturbations characterized by a nearly scale-invariant power spectrum with a small tilt. Current observations of the CMB, particularly those from the Planck satellite, support these predictions, although they have ruled out many of the simplest single-field inflation models. For example, monomial potentials such as 
$V(\phi)  \propto \phi^n$ are increasingly disfavored by data due to their predictions for the tensor-to-scalar ratio $r$, a measure of the amplitude of primordial gravitational waves, being too high compared to observations.

This observational tension has prompted theorists to explore alternative frameworks for inflation. One promising approach is the idea of scale invariance as a fundamental symmetry of nature \cite{Wetterich:2020cxq}. Scale invariance, or scale symmetry, posits that the laws of physics do not depend on the overall size or length scale, implying the absence of intrinsic mass scales. Theories that incorporate scale invariance can naturally avoid the introduction of arbitrary scales, which could require fine-tuning. This feature makes scale-invariant models attractive from a theoretical perspective, especially in the context of naturalness—the idea that physical theories should not require highly precise adjustments of parameters to agree with observations.

At the classical level, scale invariance restricts the allowed operators in the action to those that are dimensionless, significantly reducing the number of free parameters in the theory. This can lead to increased predictivity, a highly desirable feature in theoretical physics. Quantum effects, however, typically break classical scale invariance through the introduction of mass scales via renormalization. Nevertheless, certain theories retain a form of quantum scale invariance, where the quantum effective action does not contain any intrinsic dimensionful parameters. In such theories, all dimensionful quantities arise dynamically, preserving the scale-invariant structure of the theory at the quantum level.

The notion of scale invariance has been explored in various contexts in both particle physics and cosmology. In particle physics, it provides a potential resolution to the Higgs naturalness problem, where the mass of the Higgs boson is much lighter than the Planck scale, despite quantum corrections suggesting it should be much heavier. This mismatch between the observed Higgs mass and theoretical expectations raises the question of whether a symmetry, such as scale invariance, might be responsible for stabilizing the Higgs mass against large quantum corrections. Similarly, in cosmology, the observed near scale-invariance of the primordial power spectrum, along with the flatness of inflationary potentials, motivates the exploration of models where scale invariance plays a fundamental role.

In the context of inflation, the idea of scale invariance is closely related to the success of the Starobinsky model \cite{Starobinsky:1980te}, one of the most well-supported inflationary models by current observations. Starobinsky inflation introduces a modification to general relativity by adding a term quadratic in the Ricci scalar, $R^2$ to the Einstein-Hilbert action. This results in a flat inflationary potential that leads to a slow-roll phase of inflation and produces a power spectrum in excellent agreement with CMB observations. The nearly scale-invariance of the model at large curvatures, where the $R^2$-term dominates over the $R$-term, plays a crucial role in its success. More generally, models of inflation that involve modifications to gravity, such as those incorporating higher-order curvature terms, often have a scale-invariant structure.

This talk investigates a recently proposed inflationary model \cite{Rinaldi:2015uvu,Tambalo:2016eqr}. The model features a scalar field, 
$\phi$, non-minimally coupled to gravity, with an action quadratic in the Ricci scalar. This framework builds on earlier works that showed the viability of scale-invariant inflation from a theoretical standpoint, but the present study \cite{Cecchini:2024xoq} goes further by solving the full two-field dynamics of the system numerically and comparing its predictions with the latest cosmological data \cite{giare2023}, particularly from the Planck satellite and the BICEP/Keck collaborations.

\section{Model and Theoretical Background}
The model's action, written in the Jordan frame, includes a $R^2$-term, a non-minimal coupling between the scalar field and the Ricci scalar, and a quartic self-interaction for the scalar field:
\begin{equation}
S_J = \int d^4 x \, \sqrt{-g} \left( \frac{\alpha}{36} R^2 + \frac{\xi}{6} \phi^2 R - \frac{1}{2} (\partial_\mu \phi)(\partial^\mu \phi) - \frac{\lambda}{4} \phi^4 \right).
\end{equation}
Here, $\alpha$, $\xi$ and $\lambda$ are dimensionless constants, with $\alpha$ controlling the strength of the quadratic curvature term, $\xi$ controlling the non-minimal coupling of the scalar field to gravity, and $\lambda$ determining the self-interaction of the scalar field. The inclusion of the $R^2$-term is motivated by its success in Starobinsky inflation, while the non-minimal coupling and quartic self-interaction are motivated by the need to achieve scale-invariant dynamics. Indeed, in the Jordan frame, the model initially possesses classical scale symmetry, meaning that the action contains no intrinsic mass or length scale, and the laws of physics remain unchanged under the rescaling of lengths and energies. However, during the inflationary process, this symmetry is dynamically broken. The scalar field $\phi$ evolves over time, and as it stabilizes, a mass scale emerges dynamically. Specifically, this breaking of scale symmetry is associated with the scalar field reaching a non-zero value,
\begin{equation}
\langle \phi_0^2 \rangle=\frac{\xi R}{3 \lambda}
\end{equation}
which can be identified with the Planck mass
\begin{equation}
\frac{\xi}{6}\phi_0^2 R\equiv \frac{1}{2}M_{Pl}^2 R.
\end{equation}
The spontaneous breaking of scale symmetry thus leads to the generation of a mass scale, even though the initial action had no explicit mass terms. This mechanism of symmetry breaking through field dynamics is a key feature of the model, as it links the inflationary period to the generation of fundamental scales in nature, without the need for fine-tuning. Consequently, the inflaton field stabilizes the geometry of the universe, driving inflation and dynamically generating a meaningful mass scale, which ultimately governs the end of inflation and the onset of the standard cosmological evolution.

\begin{figure}
    \centering
    \includegraphics[width=8cm]{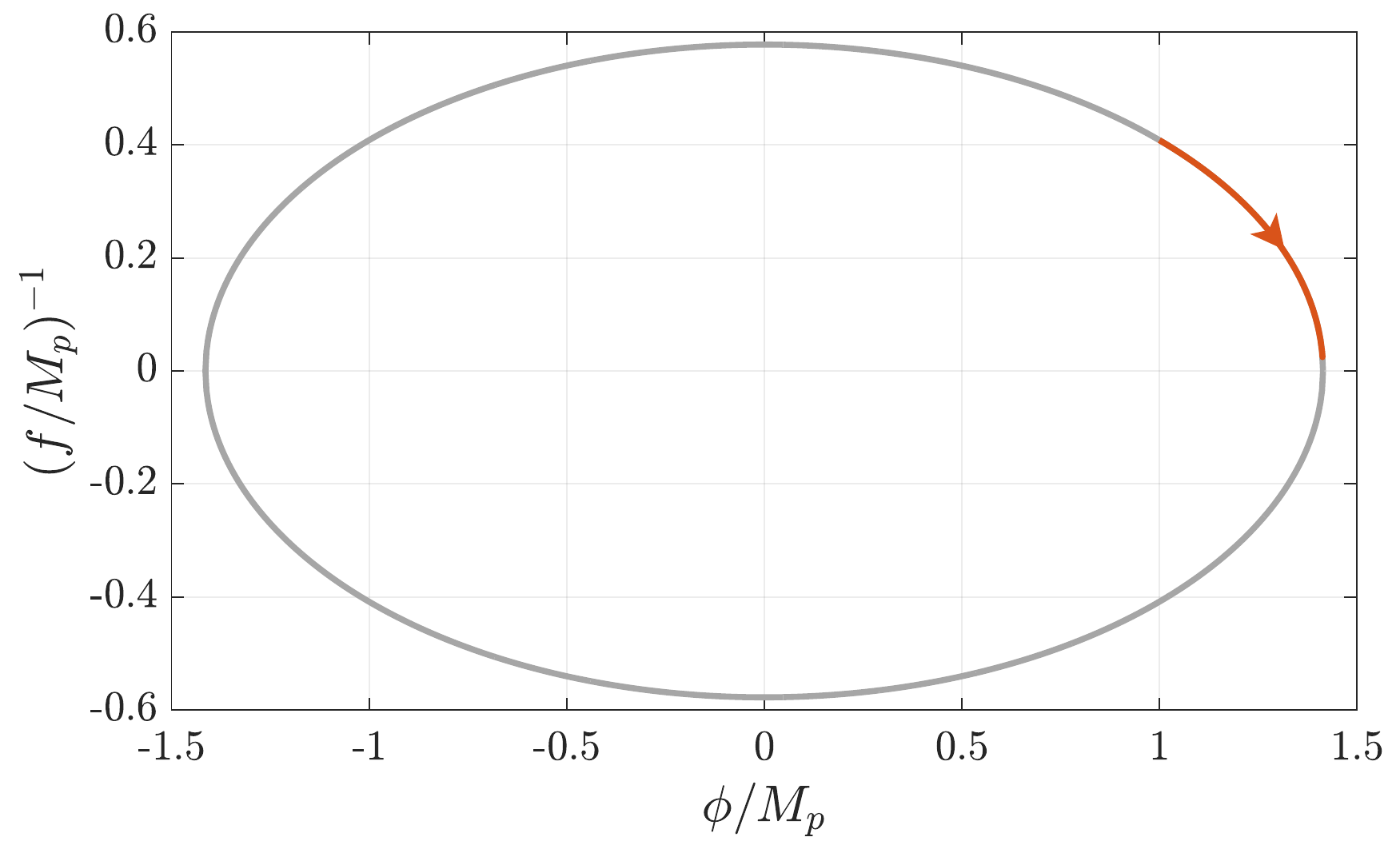}
    \caption{Analytical equation of the ellipse from the Noether current. Trajectory of the fields obtained numerically by solving the background equations (red curve). The arrow indicates
the direction of motion. Numerical integration is carried out until the end of inflation}
    \label{fig1}
\end{figure}

To analyze the model, we move from the Jordan to the Einstein frame with a conformal transformation $\tilde{g}_{\mu \nu}=e^{2 \omega(x)} g_{\mu \nu}$ \cite{}, where the gravitational sector takes the standard form of general relativity
\begin{equation}
    \mathcal{L}_E=\sqrt{-g}\biggl[ \frac{M^2}{2}R - \frac{3 M^2}{f^2}(\partial f)^2-\frac{f^2}{2M^2}(\partial \phi)^2-V(f,\phi) \biggl].
\end{equation}
In the Einstein frame, the model reduces having an effective potential for the scalar field that drives inflation. One of the key findings of the paper is that, despite involving two fields initially, the scale-invariance of the theory constrains the dynamics to follow a single-field trajectory. Indeed, the solution to the conservation equation, $\nabla_{\mu}K^{\mu} = 0$, with
\begin{equation}
K_{\mu} \equiv \partial_{\mu}K\,, \quad \quad K \equiv \dfrac{M_p^2}{2} \left ( \dfrac{\phi^2}{M_p^2} + \dfrac{6M_p^2}{\mathfrak{f}^2} \right ) \,,
\label{eq:K}
\end{equation}
reveals that the value of $K$ approaches a constant over time. This behavior dynamically breaks scale symmetry and constrains the motion of the fields along an elliptical trajectory (Fig.\ref{fig1}), ensuring that the dynamic effectively reduced to those of a single field.

As a result, the field  $f$ can be expressed in terms of $\phi$ as
\begin{equation}
    f=\frac{\sqrt{6}M_{Pl}^2}{\sqrt{2M_{Pl}^2-\phi^2}},
\end{equation}
and the Noether current helps protect the model from potential destabilization effects, such as the development of entropy perturbations, which are known to plague some of the multi-field inflationary models. 
\subsection{Field redefinition}
\begin{figure}
\quad
    \includegraphics[width=13cm]{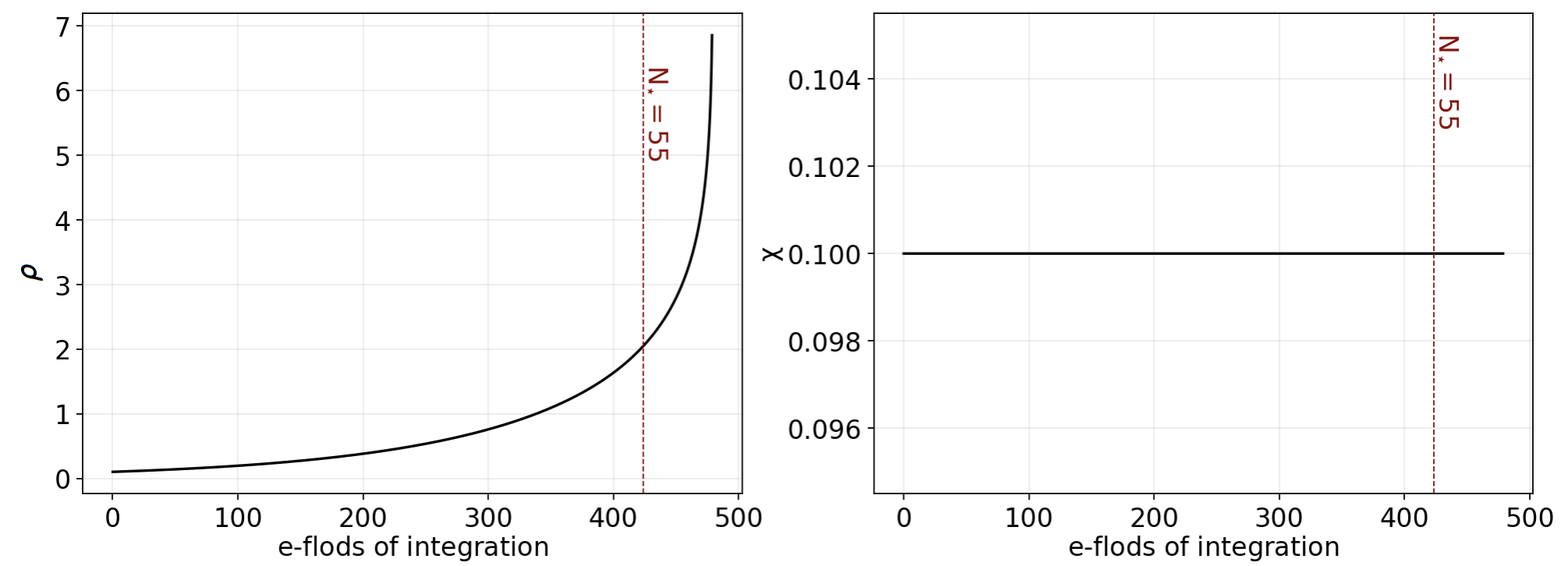}
    \caption{Trajectory of the redefined fields $\rho$ and $\chi$.}
    \label{fig:1field}
\end{figure}

\begin{figure}
\centering
    \includegraphics[width=6cm]{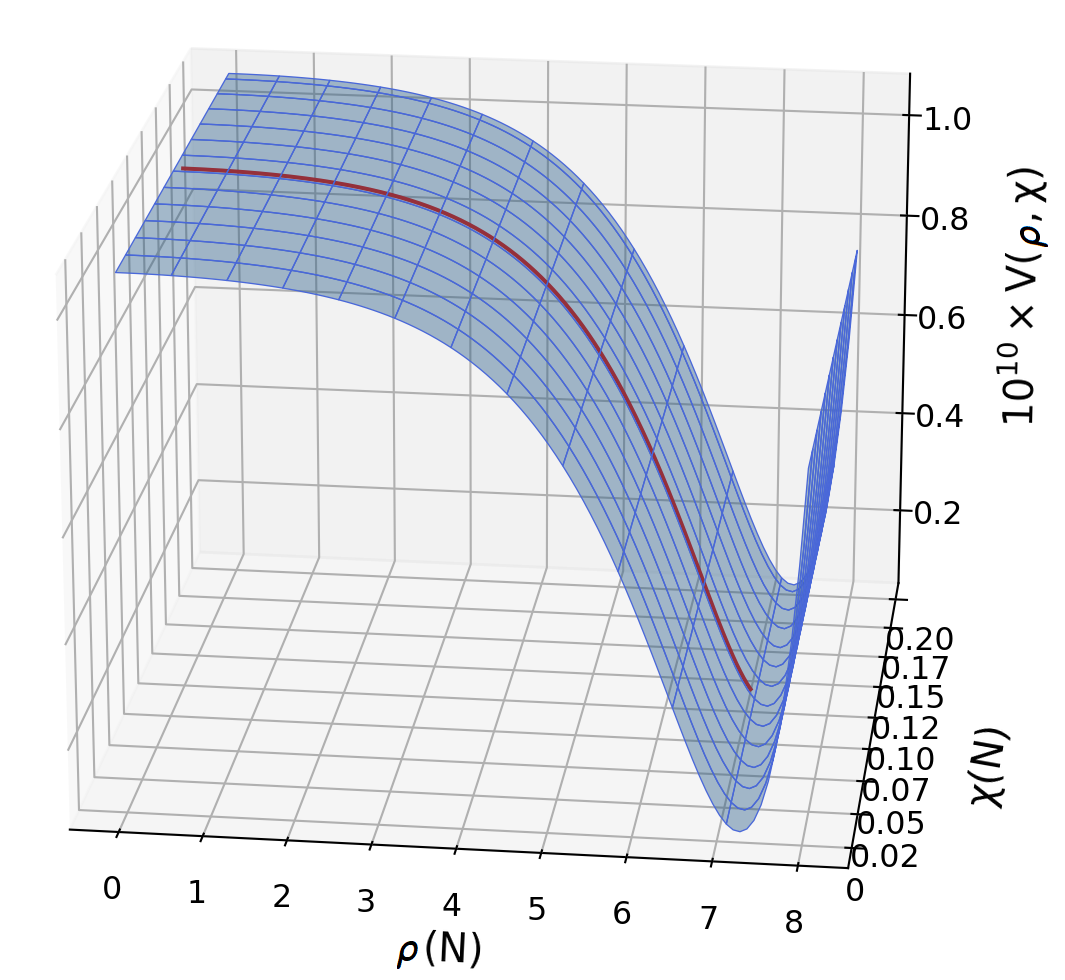}
    \caption{Trajectory of the fields on the potential \eqref{eq:potentialrhochi}.}
    \label{fig:pot}
\end{figure}

It can be better shown that by considering a field redefinition such as $\rho=\rho(f,\phi)$ and $\chi=\chi(f,\phi)$ resulting in the action of the form
\begin{equation}
    \mathcal{L}_E=\sqrt{-g}\biggl( \frac{M^2}{2}R-\frac{1}{2}\partial_{\mu}\rho \partial^{\mu}\rho-3\cosh\biggl[\frac{\rho}{\sqrt{6}M}\biggl]^2 \partial_{\mu}\chi\partial^{\mu}\chi-V(\rho) \biggl),
\end{equation}
the model can be treated in the same vein as single-field inflation,
given that its dynamical content can be shifted to one field out of the two, with the other one behaving effectively as a spectator field (Fig.\ref{fig:1field}). 
The potential in this field representation depends exclusively on $\rho$, and takes the form
\begin{eqnarray}
V(\rho) = \dfrac{9M_p^4}{4\alpha}\left[1-4\xi \sinh^2 \left ( \dfrac{\rho}{\sqrt{6}M_p} \right )+4\Omega \sinh^4 \left ( \dfrac{\rho}{\sqrt{6}M_p} \right ) \right]\,,
\label{eq:potentialrhochi}
\end{eqnarray}
where $\Omega \equiv \alpha \lambda +\xi^2$. 
It is clear that there is no need for fine-tuning as a flat plateau potential arises naturally. In this scheme, $\rho$ plays the role of the inflaton field and $\chi$ is nothing but the Goldstone boson associated to the flat directions of the potential (as shown in Fig.\ref{fig:pot}).  

\section{Inflationary predictions and observables}
We present an in-depth analysis of how the scale-invariant inflationary model fits current cosmological data, particularly from Planck and BICEP/Keck. The analysis focuses on constraining the model’s parameters using a combination of numerical simulations and Monte Carlo sampling. We utilize a custom method designed to explore the parameter space of the model efficiently.

\subsection{Methodology}
The analysis employs a Monte Carlo (MC) sampling technique to explore the model's three key parameters:
\begin{itemize}
    \item $\alpha$ controls the strength of the $R^2$ term in the Jordan frame,
    \item $\xi$ the non-minimal coupling between the scalar field $\phi$ and gravity,
    \item $\Omega$ which relates to the quartic coupling of the scalar field.
\end{itemize}
The method used is as follows:
\begin{enumerate}
    \item Initial Conditions and Parameter Ranges:
    The Monte Carlo algorithm starts by randomly selecting initial conditions for the fields and parameter values from uniform prior ranges. The prior range for the parameters is based on theoretical considerations, allowing for a comprehensive exploration of possible initial conditions.
    \item Integration of Field Equations: For each set of parameters and initial conditions, the full set of field equations is integrated numerically. The equations are solved for a maximum number of e-folds (up to $10^7$) to ensure that the model can successfully generate sufficient inflation and discard eternal inflation. The code dynamically calculates the slow-roll parameter $\epsilon$, which determines when inflation ends ($\epsilon=1$).
    \item Inflationary Observables: After ensuring that inflation ends correctly, the field dynamics during inflation are reconstructed to calculate inflationary observables such as $n_s$, $r$, $A_s$, $\alpha_s$. These quantities are computed at the moment when the pivot scale exits the horizon, typically 55 e-folds before the end of inflation.
    \item Comparison with Observational Data: The model’s predictions are compared against current cosmological data using a multi-dimensional Gaussian likelihood function. The likelihood function incorporates data from the Planck 2018 temperature and polarization measurements, as well as the latest BICEP/Keck B-mode polarization data, which constrain the tensor-to-scalar ratio $r$.
    \item Sampling the Parameter Space: Over $10^4$ points are sampled using this method, with each point weighted according to its likelihood. This large dataset allows us to derive robust constraints on the model’s parameters and test their sensitivity to changes in initial conditions.
\end{enumerate}

\subsection{Parameter Constraints}
Based on the analysis, we place stringent constraints on the model’s parameters and determine the corresponding values of inflationary observables. Here’s a breakdown of the key findings:
\begin{enumerate}
    \item Non-Minimal Coupling $(\xi)$: the analysis places a strong upper limit on the non-minimal coupling strength $\xi$. Specifically, the results exclude $\xi=1$ (the conformal coupling value) with high significance. This is a notable result, as it suggests that the model operates outside the conformal coupling regime. The constraint on $\xi$ is set to $\xi < 0.00142$ (95\% confidence level), indicating that scale-invariant inflation works well for small values of $\xi$.
    
    \item Quadratic Curvature Term ($\alpha$): the parameter $\alpha$, which controls the strength of the $R^2$-term, is constrained to a narrow range around $1.951^{+0.076}_{-0.11} \times 10^{10}$ (68\% confidence level). This result reflects the model's successful calibration against the amplitude of the primordial scalar power spectrum $A_s$. The fact that $\alpha$ is tightly constrained suggests that the model predictions for $r$ and $n_s$ are highly sensitive to this parameter, confirming the key role played by the $R^2$-term in driving inflation.
    
    \item Quartic Coupling ($\Omega$): the parameter $\Omega$, which determines the strength of the quartic self-coupling of the scalar field, is less tightly constrained compared to $\alpha$ and $\xi$. The analysis finds that it lies in the range $0.95^{+0.72}_{-2.8} \times 10^{-5}$ (68\% confidence level). The results confirm that it must be kept within specific bounds to prevent eternal inflation, where inflation would continue indefinitely and fail to transition into the hot Big Bang phase.

    \item Insensitivity to Initial Conditions: one of the key findings of the paper is that the model is insensitive to initial conditions. This means that inflation can occur for a wide range of starting field values, making the model robust and free from fine-tuning problems commonly encountered in other inflationary models.
    
\end{enumerate}

The numerical analysis also provides predictions for inflationary observables. The analysis yields a best-fit value of $n_s=0.9638^{+0.0015}_{-0.0010}$ which is consistent with the Planck 2018 result. The model predicts a lower bound on the tensor-to-scalar ratio, $r>0.00332$, which is within the sensitivity range of next-generation CMB experiments. This result is significant because the detection of primordial gravitational waves at this level would strongly support the scale-invariant inflationary model.
The running of the spectral index $\alpha_s$ which measures the scale dependence of $n_s$, is found to be small and consistent with zero, with an upper limit of $\alpha_s<1.2 \times 10^{-4}$. This result is also consistent with the Planck data.

\section{Comparison with other models}

We compare the predictions of the scale-invariant inflationary model with two well-established inflationary models: Starobinsky inflation and 
$\alpha$-attractor inflation. These comparisons are essential for understanding how the scale-invariant model fits within the broader landscape of inflationary theories and how it can be distinguished observationally from these alternatives.

\subsection{Starobinsky inflation}
Starobinsky inflation is one of the most successful inflationary models in terms of its agreement with observational data. It was initially proposed by Alexei Starobinsky in 1980 and is characterized by a modification of the standard Einstein-Hilbert action of general relativity through the addition of a term quadratic in the Ricci scalar, $R^2$. This leads to the action:
\begin{equation}
    S = \int d^4 x \, \sqrt{-g} \left( \frac{M_{Pl}^2}{2} R + \frac{\alpha}{36} R^2 \right).
\end{equation}
The scale-invariant inflationary model shares several features with Starobinsky inflation, particularly the inclusion of a quadratic $R^2$
term in the action. This leads to similar inflationary dynamics, especially when the non-minimal coupling $\xi$ in the scale-invariant model is small. In this limit, the model approaches the dynamics of Starobinsky inflation and the predictions for $n_s$ and $r$ overlap with those of Starobinsky.

\begin{figure}
\centering
\includegraphics[width=7cm]{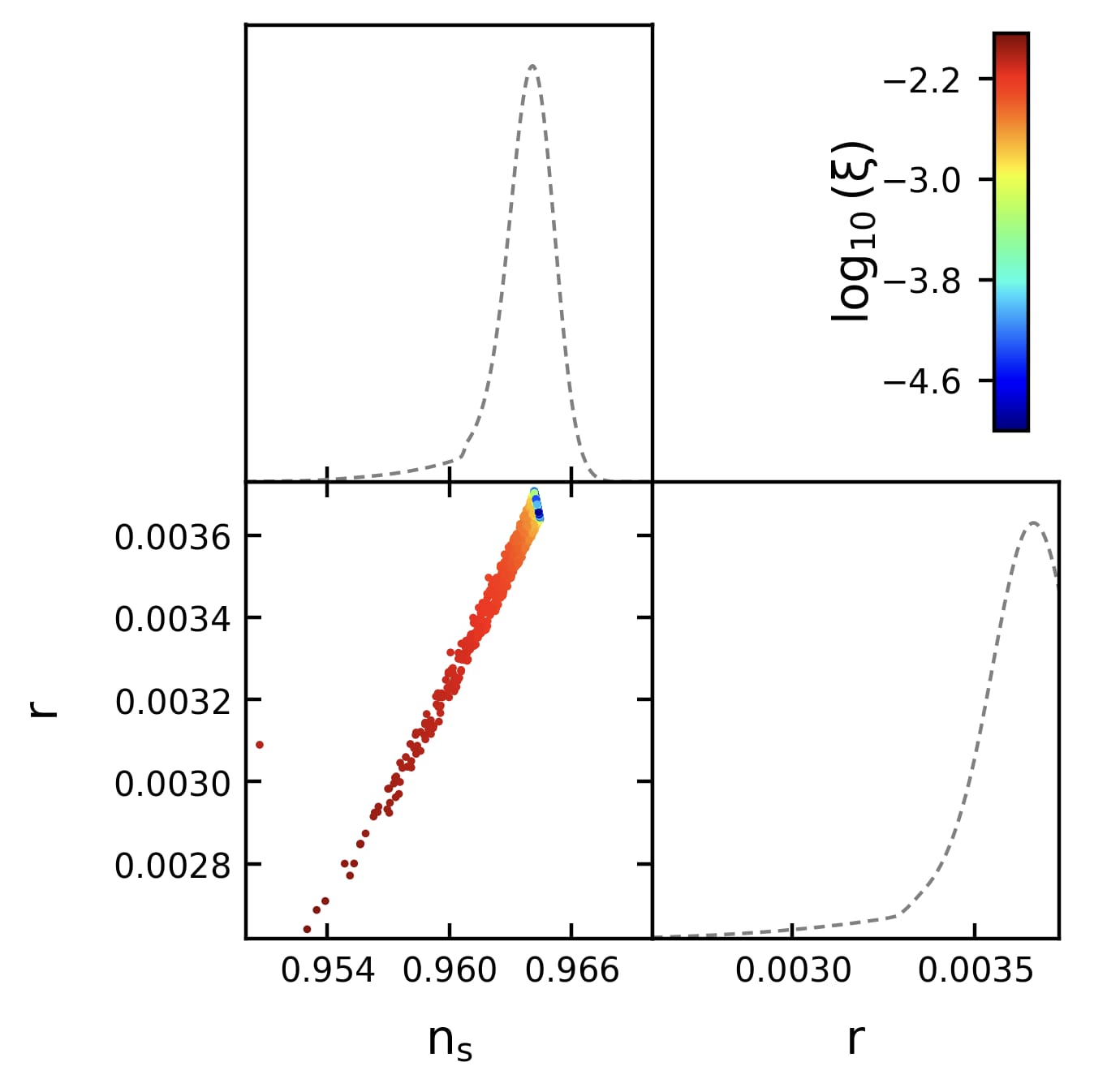}
\caption{2D scatter plot in the log10 $n_s$-$r$ plane colored by the value of $\xi$.}
\label{figpar}
\end{figure}

However, the scale-invariant model introduces a scalar field $\phi$ that is non-minimally coupled to gravity via a term proportional to $\phi^2R$. This non-minimal coupling plays a key role in distinguishing the two models (Fig.\ref{figpar}):
\begin{itemize}
    \item For small values of $\xi$: The predictions for $n_s$ and $r$ are nearly identical to those of Starobinsky inflation because the $R^2$ term dominates the dynamics. In this regime, both models predict a nearly flat inflationary potential, leading to values of $n_s \simeq 0.965$ and $r\simeq 0.0037$.
    \item For larger values of $\xi$: The scale-invariant model starts to deviate from Starobinsky inflation. As $\xi$ increases, the contribution of the non-minimal coupling becomes more significant, causing the model to predict slightly smaller values for the spectral index and the tensor-to-scalar ratio
    \begin{equation}
n_s \simeq 1-\dfrac{1}{3}\sqrt{3r + 64\xi^2}.
\end{equation}
\end{itemize}
In particular, the scale-invariant model can push the value of $r$ lower than in Starobinsky inflation, providing a potential observational distinction.

\subsection{$\alpha$-attractors}
The scale-invariant inflationary model can be seen as a specific case of the more general $\alpha$-attractor framework in certain limits. Both models incorporate scale invariance, but there are several differences.  In $\alpha$-attractors \cite{Linde:2015uga,Kallosh:2015lwa,Shojaee:2020xyr,Rinaldi:2015yoa}, the parameter $\alpha$ controls the shape of the potential and therefore affects both $n_s$ and $r$. This provides greater flexibility in fitting observational data compared to the scale-invariant model, where the non-minimal coupling $\xi$ plays a similar role. However, the scale-invariant model is more predictive, as it involves fewer free parameters than the $\alpha$-attractors.
While both models can fit current CMB data, the scale-invariant model is more constrained in terms of its predictions for $r$. For larger values of $\xi$, the scale-invariant model predicts lower values for $r$, which could help distinguish it from $\alpha$-attractors. In contrast, $\alpha$-attractors allow a broader range of possible values for $r$, depending on the choice of $\alpha$.

\section{Future observational prospects}
\begin{figure}
\centering
\includegraphics[width=7cm]{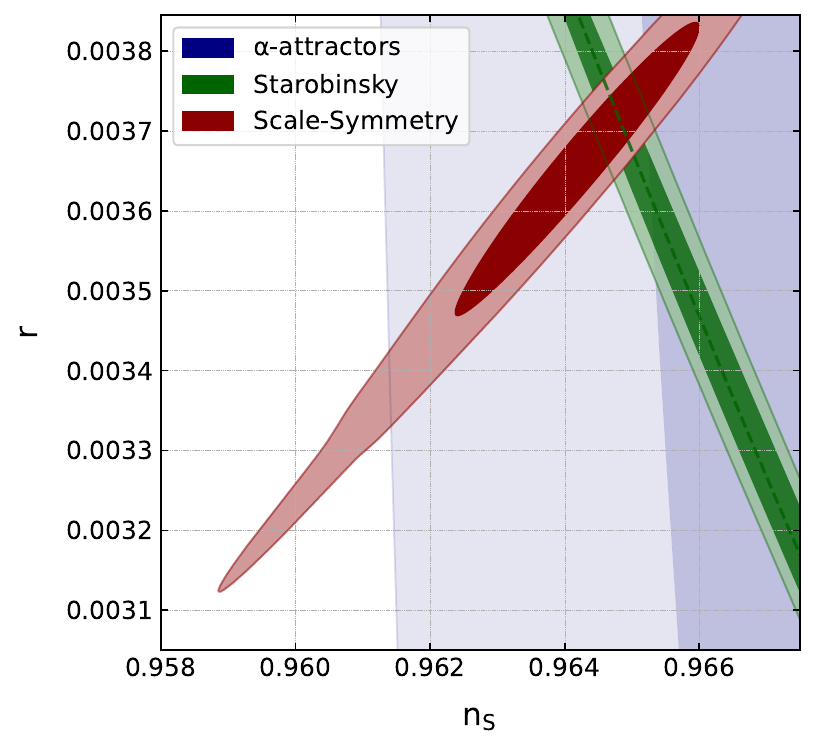}
\caption{ 2D contours in the $n_s$-$r$ plane for the scale-invariant model studied in this work (red
contours), compared to those of Starobinsky inflation (green contours) and $\alpha$-attractors (light blue
contours), obtained in light of observations from the Planck 2018 legacy release and from the BICEP/Keck collaboration (BICEP2, Keck Array and BICEP3 observations up to 2018) and sampling
procedure.}
\label{comparison}
\end{figure}

The main way to distinguish between the scale-invariant model, Starobinsky inflation, and $\alpha$-attractors lies in the tensor-to-scalar ratio $r$ (Fig.\ref{comparison}) and the potential detection of primordial gravitational waves. The detection of primordial gravitational waves, as measured by the tensor-to-scalar ratio $r$, is a key goal of next-generation CMB experiments \cite{CMB-S4:2016ple,CMB-S4:2020lpa,LiteBIRD:2022cnt}. The scale-invariant model predicts a lower bound on $r$ of approximately 
0.003, which will be within the sensitivity range of these experiments. If $r$ is detected at this level, it would provide strong support for the scale-invariant model, while ruling out models that predict much lower values for $r$. Another important distinction could come from the level of non-Gaussianity in the primordial perturbations. The scale-invariant model predicts low levels of non-Gaussianity, consistent with current observations, but future experiments might be able to probe this further.

\bibliographystyle{unsrt} 
\bibliography{ws-pro-sample}








\end{document}